\documentclass{ws-ijmpa}
\begin{document}
\markboth{In\'{e}s Cavero-Pel\'{a}ez, Kimball A. Milton, Prachi Parashar and K. V. Shajesh}
{Leading- and next-to-leading-order lateral Casimir force on corrugates surfaces}
%
\catchline{}{}{}{}{}
%
\title{LEADING- AND NEXT-TO-LEADING-ORDER LATERAL CASIMIR FORCE ON CORRUGATED SURFACES.}
\author{IN\'ES CAVERO-PEL\'AEZ}
\address{Laboratoire Kastler Brossel,Universit\'{e} Pierre et Marie Curie, ENS, CNRS,\\
Campus Jussieu, University Paris 6, Case 74, F-75252 Paris, cedex 05, France.\\
cavero@spectro.jussieu.fr}
\author{KIMBALL A. MILTON, PRACHI PARASHAR AND K. V. SHAJESH}
\address{Oklahoma Center for High Energy Physics
and Homer L. Dodge Department of Physics and Astronomy,
University of Oklahoma, Norman, OK 73019, USA.\\
milton@nhn.ou.edu, prachi@nhn.ou.edu, shajesh@nhn.ou.edu}
\maketitle
\begin{history}
\received{Day Month Year}
\revised{Day Month Year}
\end{history}
\begin{abstract}
We derive explicit analytic expressions for the lateral force for two different configurations with corrugations, parallel plates and concentric cylinders. By making use of the multiple scattering formalism, we calculate the force for a scalar field under the influence of a delta-function potential that has sinusoidal dependence in one direction simulating the corrugations. By making a perturbative expansion in the amplitude of the corrugation we find the leading order for the corrugated concentric cylinders and the next-to-leading order for the corrugated parallel plates. 
\end{abstract}
\ccode{PACS numbers: 11.25.Hf}
%
%
\section{Energy and Green's functions. }
We consider a massless scalar field $\psi$ interacting with a cylindrical $\delta$-function potential background with corrugations,
\begin{equation}
\mathcal{L}_{int}=-\frac{1}{2}\lambda_1\delta(r-a_1-h_1(\theta))-\frac{1}{2}\lambda_2\delta(r-a_2-h_2(\theta))
\end{equation}
where indices 1 and 2 represent the two concentric cylinders with radii $a_1$ and $a_2$ shown on the left side of Fig.~\ref{corru}.
The functions $h_1(\theta)$ and $h_2(\theta)$ describe the corrugations associated with the cylinders and we are going to assume they are sinusoidal.
Due to the geometry of our set up we have translational invariance on the $z$-axes as well as time independence allowing us to make a Fourier transform of the Green's function on those coordinates,
\begin{equation}
G(x,x^\prime)
= \int \frac{d\omega}{2\pi} \,e^{-i\omega (t - t^\prime)}
  \int \frac{dk}{2\pi} \,e^{i k (z - z^\prime)}g(r,r^\prime,\theta,\theta^\prime;k,\omega),\end{equation}
where the reduced Green's function satisfies the differential equation,
\begin{eqnarray}
\bigg [-\frac{1}{r}\frac{\partial}{\partial r}r\frac{\partial}{\partial r}-\frac{1}{r^2}\frac{\partial^2}{\partial\theta^2}-\omega^2+k^2-\lambda_1\delta(r-a_1-h_1(\theta))\nonumber\\
-\lambda_2\delta(r-a_2-h_2(\theta))\bigg ]
g(r,r^\prime,\theta,\theta^\prime;k,\omega)
= \frac{\delta (r-r^\prime)}{r}\delta(\theta-\theta^\prime).
\label{g-differential-eq}
\end{eqnarray}
The central formula in our calculation is the multiple scattering formula for the Casimir energy \cite{kenneth2006,emig2008,milton2008}, 
\begin{equation}
\Delta E = \frac{i}{2 \tau} \,\rm{Tr} \,\rm{ln} \,G {G^{(0)}}^{-1},
\label{CEoriginal}
\end{equation}
where $G^{(0)}$ is the Green's function associated with the background. In this sense we are going to define the background as two concentric semitransparent cylinders and the corrugations are considered to be a perturbation over such a background. Therefore $G^{(0)}=\int\frac{d\omega}{2\pi}e^{-i\omega (t - t^\prime)}\int\frac{dk}{2\pi}e^{i k (z - z^\prime)}\sum_{m=-\infty}^{\infty} \frac{1}{2\pi}e^{im(\theta - \theta^\prime)}g_m^{(0)}(r,r^\prime;k,\omega)$ satisfies 
\begin{eqnarray}
\bigg [-\frac{1}{r}\frac{\partial}{\partial r}r\frac{\partial}{\partial r}-\omega^2+k^2+\frac{m^2}{r^2}+V_1^{(0)}(r)+V_2^{(0)}(r)\bigg ]
g_m^{(0)}(r,r^\prime;k,\omega)
= \frac{\delta (r-r^\prime)}{r}.
\label{g0-differential-eq}
\end{eqnarray}
Here $V_i^{(0)}(r)=\lambda_i\delta(r-a_i)$ is the potential for the background associated with cylinder $i$ and we define $\Delta V_i$ as the deviation of the total potential $V_i(r,\theta)=\lambda_i\delta(r-a_i-h_i(\theta))$ from it, 
\begin{equation}
\Delta V_i(r,\theta)=V_i(r,\theta)-V_i^{(0)}(r)
= \sum_{n=1}^\infty \frac{[-h_i(\theta)]^n}{n!}
\frac{\partial^n}{\partial r^n} V_i^{(0)}(r)= \sum_{n=1}^\infty V_i^{(n)}(r,\theta).
\label{potential-expansion}
\end{equation}
This allows us to extract the interaction term $E_{12}$ from Eq.~(\ref{CEoriginal}), which is the only term in $\Delta E$ that gives rise to the torque. It is expressed in terms of the perturbation potential and the Green's function $G_i$ associated to each cylinder,
\begin{equation}
E_{12} = - \frac{i}{2\tau} \,\rm{Tr} \,\rm{ln}
\Big[ 1 - G_1 \Delta V_1 G_2 \Delta V_2 \Big].
\label{CEcorrug}
\end{equation}
where $G_i$ satisfies $\left[-\partial^2 + V_1^{(0)}(r) + V_2^{(0)}(r) + \Delta V_i(r,\theta)\right]G_i(r,\theta) = 1$, and can be expanded in terms of the Green's function associated with the background, 
\begin{equation}
G_i(r,\theta)=G^{(0)}-G^{(0)}\Delta V_i(r,\theta)G^{(0)}+G^{(0)}\Delta V_i(r,\theta)G^{(0)}\Delta V_i(r,\theta)G^{(0)}-...,
\end{equation}
for $i=1, 2$. Once we know the energy associated with this configuration, we can calculate the torque corresponding to a shift of one of the cylinders with respect to the other described by an angular rotation $\theta_0$\footnote{For the case of the parallel plates, we calculate the equivalent lateral force $F_{lat}=-\frac{\partial E_{12}}{\partial y_0}$, where $y_0$, is the lateral shift between the plates.}
\begin{equation}
{\cal T} = - \frac{\partial E_{12}}{\partial \theta_0}.
\label{torE}
\end{equation}
%
%
\section{Leading order contribution. Corrugated concentric cylinders}
We concentrate now in the case when the corrugations can be treated as small compared to their wavelength. In this case we can make an expansion in the corrugation amplitude in Eq.~(\ref{CEcorrug}) and keep the leading order to start with.

Let's assume sinusoidal corrugations and induce a rotation by an angle $\theta_0$ on one of the cylinders. Then these corrugations can be expressed as, $h_1(\theta)= h_1 \sin [\nu_1(\theta + \theta_0)]$ and $h_2(\theta)= h_2 \sin [\nu_2 \theta]$,
where $h_{1,2}$ are the amplitudes and $\nu$ is the frequency associated with the corrugations. The leading order contribution plays a role in the total energy only when both cylinders have the same frequency, $\nu_1=\nu_2=\nu$, giving rise to a second order term in the amplitude,
\begin{equation}
E_{12}^{(2)} = \frac{i}{2\tau} \,\rm{Tr}
\Big[ G^{(0)} V_1^{(1)} G^{(0)} V_2^{(1)} \Big]= \frac{i}{2\tau} \,\rm{Tr}
\mathcal{K}^{(2)}.
\label{E12(2)-general}
\end{equation}
In terms of the reduced Green's functions, the $\rm{Tr}\,\mathcal{K}^{(2)}$ is given by:
\begin{eqnarray}
\rm{Tr}\mathcal{K}^{(2)}&=&\frac{\lambda_1\lambda_2}{4\pi^2}\int\frac{d\omega dk}{(2\pi)^2}e^{-i\omega (t - t^\prime)}e^{i k (z - z^\prime)}\int d\theta d\theta^\prime\int rdr \,r^\prime dr^\prime \sum_{m,m^\prime=-\infty}^{\infty}h_1(\theta^\prime)h_2(\theta)\nonumber\\
&\times& e^{i\theta(m-m^\prime)}e^{i\theta^\prime (m^\prime -m)}g_m^{(0)}(r,r^\prime)\Big[\frac{\partial}{\partial r^\prime}\delta(r^\prime-a_1)\Big]g_{m^\prime}^{(0)}(r^\prime,r)\Big[\frac{\partial}{\partial r}\delta(r-a_2)\Big],
\end{eqnarray}
where we have used Eq.~(\ref{potential-expansion}). To evaluate this expression we integrate by parts in $r$ and $r^\prime$ and Fourier transform the functions $h_i(\theta)$ that describe the corrugations on the cylinders, $h_i(\theta)=\sum_{\nu=-\infty}^{\infty}\frac{1}{2\pi}e^{\nu\theta}\tilde h_\nu$. We can now integrate on $\theta$ and switch to imaginary frequencies by an Euclidean rotation, $\omega\rightarrow i\zeta$,
\begin{eqnarray}
\frac{E_{12}^{(2)}}{L_z} &=& -\frac{\lambda_1\lambda_2}{(16\pi)^3} 
\sum_{m=-\infty}^{\infty}\sum_{m^\prime=-\infty}^{\infty}
(\tilde{h}_1)_{m-m^\prime} (\tilde{h}_2)_{m^\prime-m}\int_0^\infty \kappa d\kappa \frac{\partial}{\partial r} \,\frac{\partial}{\partial \bar{r}}\nonumber\\
&&\times\Big[ \,r \,\bar{r} \,g_m^{(0)}(r,\bar{r};\kappa) 
       \,g_{m^\prime}^{(0)}(\bar{r},r;\kappa) \Big]\bigg|_{\bar{r}=a_1, r=a_2}.
\label{E12(2)}
\end{eqnarray}
Here $\kappa^2 = k_z^2 -\omega^2=k_z^2 + \zeta^2$ and the $g_m^{(0)}$'s are the solutions of Eq.~(\ref{g0-differential-eq}) in terms of the modified Bessel functions. The explicit form of the  $g_m^{(0)}$'s in the different regions are sorted out in Ref.~\refcite{gearsII} and they have been used there to evaluate the integral in the above equation. We denote this integral by $I_{mm^\prime}^{(2)}$\footnote{The derivatives of the Green's functions are calculated following the prescription described in appendix A of Ref.~\refcite{gearsI}.},
\begin{eqnarray}
I_{mm^\prime}^{(2)}(a_1,a_2;\kappa)
&=& \frac{\lambda_1 \lambda_2}{\Delta \tilde{\Delta}}
\bigg[\kappa a_1I_1 K_2\Big( \tilde{I}_1^\prime \tilde{K}_2+\frac{\lambda_1}{2\kappa}\tilde{I}_1 \tilde{K}_2 \Big)
+ \kappa a_1 \, 
\Big( I_1^\prime K_2 + \frac{\lambda_1}{2\kappa}I_1 K_2 \Big)\tilde{I}_1 \tilde{K}_2
\nonumber\\
&&+\kappa a_2 I_1 K_2\Big(\tilde{I}_1 \tilde{K}_2^\prime 
  - \frac{\lambda_2}{2\kappa}\tilde{I}_1 \tilde{K}_2 \Big)
+ \kappa a_2\Big( I_1 K_2^\prime - \frac{\lambda_2}{2\kappa}I_1 K_2 \Big)\tilde{I}_1 \tilde{K}_2\nonumber\\
&&+ I_1 K_2\tilde{I}_1 \tilde{K}_2+ \kappa a_1 \kappa a_2
\Big( I_1 K_2^\prime - \frac{\lambda_2}{2\kappa} \,I_1 K_2 \Big)
\Big( \tilde{I}_1^\prime \tilde{K}_2 
      + \frac{\lambda_2}{2\kappa} \,\tilde{I}_1 \tilde{K}_2 \Big)\nonumber\\
&&+ \kappa a_1 \kappa a_2\Big( I_1^\prime K_2 + \frac{\lambda_1}{2\kappa} \,I_1 K_2 \Big)
\Big( \tilde{I}_1 \tilde{K}_2^\prime
      - \frac{\lambda_2}{2\kappa} \,\tilde{I}_1 \tilde{K}_2 \Big)\nonumber
\end{eqnarray}
\begin{eqnarray}
&&+\kappa a_1 \kappa a_2
\Big( I_1^\prime K_2^\prime
+ \frac{\lambda_1}{2 \kappa} \,I_1 K_2^\prime
- \frac{\lambda_2}{2 \kappa} \,I_1^\prime K_2-\frac{\lambda_1}{2 \kappa} \frac{\lambda_2}{2 \kappa} \,I_1 K_2 \Big)
\tilde{I}_1 \tilde{K}_2\nonumber\\
&&+\kappa a_1 \kappa a_2 \,I_1 K_2
\Big( \tilde{I}_1^\prime \tilde{K}_2^\prime
+ \frac{\lambda_1}{2 \kappa} \,\tilde{I}_1 \tilde{K}_2^\prime
- \frac{\lambda_2}{2 \kappa} \,\tilde{I}_1^\prime \tilde{K}_2-\frac{\lambda_1}{2 \kappa} \frac{\lambda_2}{2 \kappa} 
  \,\tilde{I}_1 \tilde{K}_2 \Big)
\bigg].
\label{I2cyl}
\end{eqnarray}
where $\Delta = 1 + \lambda_1 a_1 \, I_1 K_1 + \lambda_2 a_2 \, I_2 K_2 
+ \lambda_1 a_1 \lambda_2 a_2 \, I_1 K_2 \, \big( I_2 K_1 - I_1 K_2 \big)$ and we have used the notation $I_{1,2} \equiv I_m(\kappa a_{1,2})$ (same for $K$)
and with index $m^\prime$ for the tilde's.

On the other hand the Fourier transforms, $(\tilde h_i)_{mm^\prime}$ can be written explicitly as
\begin{equation}
(\tilde{h}_1)_m = h_1 \frac{\pi}{i}\Big[ \,e^{i\nu\theta_0} \delta_{m,\nu}
- e^{-i\nu\theta_0} \delta_{m,-\nu} \Big],\qquad
(\tilde{h}_2)_m = h_2 \frac{\pi}{i}\Big[\delta_{m,\nu}-\delta_{m,-\nu} \Big],
\end{equation}
and Eq.~(\ref{E12(2)}) becomes $\frac{E_{12}^{(2)}}{L_z} = - \cos(\nu\theta_0) \,\frac{h_1 h_2}{8\pi}
\sum_{m=-\infty}^{\infty} \int_0^\infty \kappa \,d\kappa
\, I^{(2)}_{m,m+\nu}(a_1,a_2;\kappa)$.

\begin{romanlist}[(ii)]
\item Dirichlet limit.
For the case of the Dirichlet limit ($a \lambda_{1,2} \gg 1$), Eq.~\eqref{I2cyl} takes the form
\begin{eqnarray}
I_{mm^{\prime}}^{(2)D}(a_1,a_2;\kappa)&=& - \frac{1}{a_1 a_2} \frac{1}{[I_2 K_1 - I_1 K_2]}\frac{1}{[\tilde{I}_2 \tilde{K}_1 - \tilde{I}_1 \tilde{K}_2]}.
\end{eqnarray}
If we plug this into the above equation for $E_{12}^{(2)}$ and use Eq.~\eqref{torE}, we find
\begin{equation}
\frac{{\cal T}^{(2)D}}{2\pi R\,L_z}
=- \nu \sin(\nu\theta_0) \,\frac{h_1h_2}{16 \pi^2 R}\sum_{m=-\infty}^{\infty}\int_0^\infty \kappa d\kappa\,I_{mm+\nu}^D(a_,a_2;\kappa),
\end{equation}
where we have divided by a factor of $2\pi R$, which is the 
mean circumference.
\item Weak coupling limit. For the case of weak coupling ($a \lambda_{1,2} \ll 1$)  the Casimir torque per unit area becomes
\begin{equation}
\frac{{\cal T}^{(2)W}}{2\pi R\,L_z}
= \nu \sin(\nu\theta_0) \, \frac{\lambda_1 \lambda_2}{32 \pi^2\,a}
\,\frac{h_1}{a} \frac{h_2}{a}\frac{\alpha^3}{2} \frac{\partial}{\partial \alpha}
\bigg[ \frac{1}{\alpha^2} \left( \frac{1-\alpha}{1+\alpha} \right)^\nu
(1-\alpha^2) (1 + 2\alpha\nu + \alpha^2) \bigg],
\end{equation}
where $\alpha=\frac{a}{2R}$
\end{romanlist}
%
%
\section{Next-to-leading-order contribution. Parallel corrugated plates.}
\begin{figure}
\begin{tabular}{cc}
\includegraphics[width=50mm]{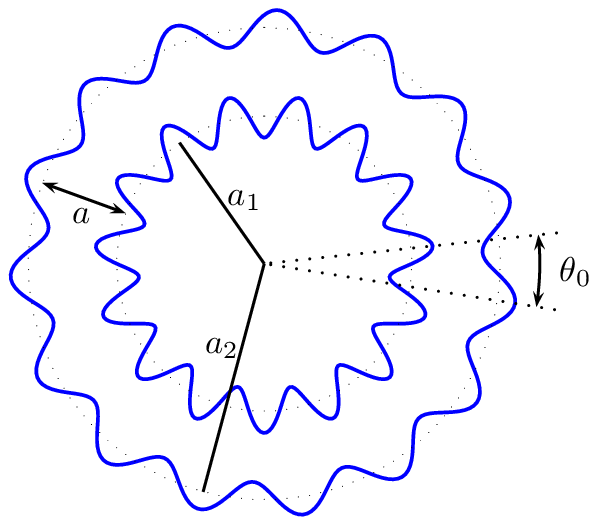}&
\includegraphics[width=70mm]{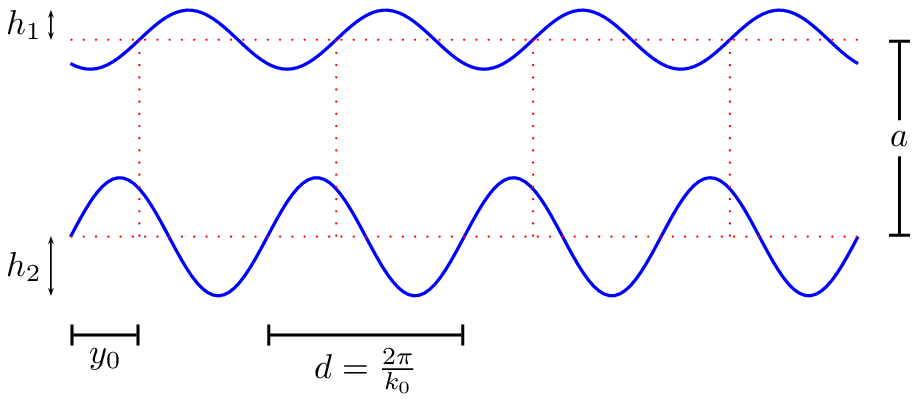}\\
\end{tabular}
\caption{Non-contact gears: On the left, concentric corrugated cylinders 
with the same corrugation frequency, $\nu=15$, on each cylinder. 
$\theta_0$ is the angular shift between the gears. On the right, parallel plates with sinusoidal corrugations. The lateral shift between the plates is $y_0$.}
\label{corru}
\end{figure}

The calculation of the leading-order contribution in the case of parallel plates with corrugations, is analogous to the case of concentric cylinders and we are not going to give its derivation here\footnote{A derivation of the leading order following the method described here can be found in Ref.~\refcite{gearsI}. The electromagnetic case at the same order has been calculated in Ref.~\refcite{emig2003,rodrigues2006}.}. In the case of parallel plates, the corrugations are described by the functions $h_1(y)=h_1\sin [k_{0}(y + y_0)]$ and $h_2(y)=h_2 \sin [k_{0} y],$, where $k_0=2\pi/d$ is the wavenumber corresponding to the corrugation wavelength $d$ (see Fig.~\ref{corru}), and $y_0$ is the shift that gives rise to the lateral force.
The next-to-leading-order contribution is calculated from Eq.~(\ref{CEcorrug}) by keeping terms up to fourth order in the corrugation amplitudes. The reason is that, as happens for corrugated concentric cylinders, for the case of corrugations with the same wavelength on both plates, the leading order is a second-order contribution while the third-order perturbation terms do not contribute to the energy. Therefore, the next-to-leading order is a fourth order term.  Following the notation introduced in Eq.~(\ref{E12(2)-general}) we can write $E_{12}^{(4)}=\frac{i}{2\tau} \,\rm{Tr}\,
\mathcal{K}^{(4)}$, where in a simplified notation $\mathcal{K}^{(4)}$ can now be written as,
\begin{eqnarray}
\mathcal{K}^{(4)}&=&\Big[G^{(0)}V_1^{(2)}G^{(0)}V_2^{(2)}-G^{(0)}V_1^{(1)}G^{(0)}V_1^{(1)}G^{(0)}V_2^{(2)}-G^{(0)}V_1^{(1)}G^{(0)}V_1^{(2)}G^{(0)}V_2^{(1)}\nonumber\\
&+&G^{(0)}V_1^{(1)}G^{(0)}V_1^{(1)}G^{(0)}V_2^{(1)}G^{(0)}V_2^{(1)}+\frac{1}{2}G^{(0)}V_1^{(1)}G^{(0)}V_2^{(1)}G^{(0)}V_1^{(1)}G^{(0)}V_2^{(1)}\Big]\nonumber\\
&+&\Big[G^{(0)}V_1^{(1)}G^{(0)}V_2^{(3)}-G^{(0)}V_1^{(1)}G^{(0)}V_1^{(1)}G^{(0)}V_2^{(2)}-G^{(0)}V_1^{(1)}G^{(0)}V_2^{(2)}G^{(0)}V_2^{(1)}\nonumber\\
&+&G^{(0)}V_1^{(1)}G^{(0)}V_2^{(1)}G^{(0)}V_2^{(1)}G^{(0)}V_2^{(1)}\Big]\nonumber\\
&+&\Big[G^{(0)}V_1^{(3)}G^{(0)}V_2^{(1)}-G^{(0)}V_1^{(2)}G^{(0)}V_2^{(1)}G^{(0)}V_2^{(1)}-G^{(0)}V_1^{(2)}G^{(0)}V_1^{(1)}G^{(0)}V_2^{(1)}\nonumber\\
&+&G^{(0)}V_1^{(1)}G^{(0)}V_1^{(1)}G^{(0)}V_1^{(1)}G^{(0)}V_2^{(1)}\Big]=\mathcal{K}^{(2,2)}+\mathcal{K}^{(1,3)}+\mathcal{K}^{(3,1)}.
\label{k4th}
\end{eqnarray}
The upper-indices in the last line indicate the order of contribution of the potentials $V_1$ and $V_2$ respectively. We explicitly write one of the above terms to illustrate the procedure. $\mathcal{K}^{(2,2)}$ can be written as $\mathcal{K}^{(2,2)A}+\frac{1}{2}\mathcal{K}^{(2,2)B}$ where the last term gives the following contribution to the interaction energy,
\begin{eqnarray}
\frac{E_{12}^{(2,2)B}}{L_x}
&=& \frac{-\lambda_1^2\lambda_2^2}{4\pi}\int\frac{dk_1}{2\pi}\frac{dk_2}{2\pi} 
\frac{dk_3}{2\pi}\frac{dk_4}{2\pi}
\,\tilde{h}_1(k_1-k_2) \tilde{h}_2(k_2-k_3) 
\tilde{h}_1(k_3-k_4) \tilde{h}_2(k_4-k_1)\nonumber\\
&\times&\int_0^\infty \bar\kappa d\bar\kappa\frac{\partial}{\partial z_1}\frac{\partial}{\partial z_2}\frac{\partial}{\partial z_1}\frac{\partial}{\partial z_2}g^{(0)}(z_1,z_2)g^{(0)}(z_2,z_3)g^{(0)}(z_3,z_4)g^{(0)}(z_4,z_1).
\label{E12(22)B}
\end{eqnarray}
Here, similarly to the case of concentric cylinders $\tilde h_1(k)=h_1\frac{\pi}{i}\Big[ e^{ik_0y_0} \delta (k-k_0) - e^{-ik_0y_0} \delta (k+k_0) \Big]$ and we get $\tilde h_2(k)$ from  $\tilde h_1(k)$ by setting $y_0=0$\footnote{The reduced Green's functions used here satisfy the Cartesian coordinate version of Eq.~(\ref{g-differential-eq}) and the solutions can be found in appendix A of Ref.~\refcite{gearsI}.}. The rest of the terms in  Eq.~(\ref{k4th}) can be written n the same fashion.
\begin{romanlist}[(ii)]
\item Dirichlet limit. The next-to-leading order contribution to the lateral Casimir force 
in the limit $(a\lambda_{1,2}\gg 1)$ becomes
\vspace{-.3cm}
\begin{eqnarray}
F_\text{Lat,D}^{(4)}
&=&2 \,k_0a \, \sin (k_0y_0) \,\left|F_\text{Ca's,D}^{(0)} \right| 
\,\frac{h_1}{a} \frac{h_2}{a} \,\frac{15}{4}
\frac{1}{\pi^4} \int_0^\infty \bar{s}\,d\bar{s} 
\int_{-\infty}^{\infty} dt\Bigg{\{}\frac{1}{2}\left( \frac{h_1^2}{a^2} + \frac{h_2^2}{a^2} \right)\nonumber\\
&&\times\frac{s}{\sinh s} \frac{s_+}{\sinh s_+}
\left[ 4 \frac{s}{\tanh s} \frac{s_-}{\tanh s_-}
+ 2 \frac{s}{\tanh s} \frac{s_+}{\tanh s_+} - s^2 - s_-^2 \right]\\
&&- 2 \cos (k_0y_0) \,\frac{h_1}{a} \frac{h_2}{a}\left[ \frac{s^2}{\sinh^2 s} \frac{s_-^2}{\sinh^2 s_-}
+ 2 \frac{s^2}{\tanh^2 s} \frac{s_+}{\sinh s_+} \frac{s_-}{\sinh s_-}\nonumber
\right]\Bigg{\}},
\label{latf-4D}
\end{eqnarray}
where $s^2 = \bar{s}^2 + (ka)^2$, and
$s^2_\pm = \bar{s}^2 + (ka \pm k_0a)^2$.
\item Weak coupling limit. If we now take the limit $a\lambda_{1,2}\ll 1$ we find that the fourth-order
contribution to the lateral Casimir force in the weak coupling equals
\begin{eqnarray}
\vspace{-.3cm}
F_\text{Lat,W}^{(4)}
&=& k_0a \, \sin (k_0y_0) \,\left|F_\text{Cas,W}^{(0)} \right|
\,\frac{h_1}{a} \frac{h_2}{a} \,\frac{3}{2}
\Bigg[ \left( \frac{h_1^2}{a^2} + \frac{h_2^2}{a^2} \right)e^{-k_0a} \sum_{m=0}^4 \, \frac{(k_0a)^m}{m!} \nonumber\\
&&- 2 \cos (k_0y_0) \,\frac{h_1}{a} \frac{h_2}{a}e^{-2k_0a} \sum_{m=0}^4 \, \frac{(2k_0a)^m}{m!}\Bigg].
\label{latf-4W}
\end{eqnarray}
\end{romanlist}
%
%
\vspace{-.95cm}
\section{Conclusions}
We have used multiple scattering techniques to calculate the leading-order torque between concentric cylinders with corrugations and the next-to-leading-order corrugated parallel plates. Our results for the Casimir torque on corrugated cylinders reproduce
the results for the lateral force on corrugated parallel plates
in the limit of large radii and small corrugation wavelengths, see Ref.~\refcite{gearsII} and \emph{Lateral Casimir forces on parallel plates and concentric cylinders with corrugations} by the same authors in the procceedings of the seminar ``60 years of the Casimir Effect''.
%
%
\vspace{-.4cm}
\section*{Acknowledgments}
We thank the US National Science Foundation (Grant No. PHY-0554926)
and the US Department of Energy (Grant No. DE-FG02-04ER41305)
for partially funding this research.
ICP would like to thank the French National Research Agency (ANR)
for support through Carnot funding. We especially thank the organizers of the Alexander Friedmann seminar and of the Satellite Symposium devoted to 60 years of the Casimir Effect.


\vspace{-.4cm}
\begin{thebibliography}{99}
\bibitem{kenneth2006}
  O. Kenneth and I. Klich, {\it Phys. Rev. Lett.} {\bf 97}, 160401 (2006).
\bibitem{emig2008}
  T.~Emig, N.~Graham, R.~L.~Jaffe and M.~Kardar,
  {\it Phys.\ Rev.\  D} {\bf 77}, 025005 (2008)
\bibitem{milton2008}
  K.~A.~Milton and J.~Wagner,
  {\it J.\ Phys.\ A } {\bf 41}, 155402 (2008)
\bibitem{gearsI}
  I. Cavero-Pel\'aez, K. A.  Milton, P. Parashar and K. V. Shajesh, 
  {\it Phys.\ Rev.\  D} {\bf 78}, 065018 (2008)
\bibitem{emig2003}
  T. Emig, A. Hanke, R. Golestanian and M. Kardar, {\it Phys. Rev. A} {\bf 67}, 022114 (2003).
\bibitem{rodrigues2006}
  R.~B.~Rodrigues, P.~A.~Maia~Neto, A.~Lambrecht, and S.~Reynaud,
  {\it Phys. Rev. Lett.} {\bf 96}, 100402 (2006).
\bibitem{gearsII}
  I. Cavero-Pel\'aez, K. A.  Milton, P. Parashar and K. V. Shajesh, 
  {\it Phys.\ Rev.\  D} {\bf 78}, 065019 (2008)
\end{thebibliography}
\end{document}